\documentclass[12pt,preprint]{aastex}
\usepackage{graphicx,natbib,amsmath}

\newcommand{\lapprox} {\, \lower3pt\hbox{$\sim$}\llap{\raise2pt\hbox{$<$}}\,}
\newcommand{\gapprox} {\, \lower3pt\hbox{$\sim$}\llap{\raise2pt\hbox{$>$}}\,}

\begin{document}

\title{SUPPRESSION OF PARALLEL TRANSPORT IN TURBULENT MAGNETIZED PLASMAS AND ITS IMPACT ON NON-THERMAL AND THERMAL ASPECTS OF SOLAR FLARES}

\author{Nicolas H. Bian\altaffilmark{1},
        Eduard P. Kontar\altaffilmark{1}, and A. Gordon Emslie\altaffilmark{2}}

\altaffiltext{1}{School of Physics \& Astronomy, University of Glasgow, Glasgow G12 8QQ, Scotland, UK \\
(n.bian@physics.gla.ac.uk, eduard.kontar@astro.gla.ac.uk)}

\altaffiltext{2}{Department of Physics \& Astronomy, Western Kentucky University, Bowling Green, KY 42101 (emslieg@wku.edu)}

\begin{abstract}
The transport of the energy contained in electrons, both thermal and suprathermal, in solar flares plays a key role in our understanding of many aspects of the flare phenomenon, from the spatial distribution of hard X-ray emission to global energetics. Motivated by recent {\em RHESSI} observations that point to the existence of a mechanism that confines electrons to the coronal parts of flare loops more effectively than Coulomb collisions, we here consider the impact of pitch-angle scattering off turbulent magnetic fluctuations on the parallel transport of electrons in flaring coronal loops. It is shown that the presence of such a scattering mechanism in addition to Coulomb collisional scattering can significantly reduce the parallel thermal and electrical conductivities relative to their collisional values.  We provide illustrative expressions for the resulting thermoelectric coefficients that relate the thermal flux and electrical current density to the temperature gradient and the applied electric field. We then evaluate the effect of these modified transport coefficients on the flare coronal temperature that can be attained, on the post-impulsive-phase cooling of heated coronal plasma, and on the importance of the beam-neutralizing return current on both ambient heating and the energy loss rate of accelerated electrons.  We also discuss the possible ways in which anomalous transport processes have an impact on the required overall energy associated with accelerated electrons in solar flares.
\end{abstract}


\section{Introduction}

A solar flare involves a complex set of energy release and transport mechanisms, with both non-thermal and thermal elements \citep[see, e.g.,][for reviews]{1988psf..book.....T,2011SSRv..159..107H}.
In particular, it is generally accepted \citep[e.g.,][]{2011SSRv..159..301K} that a significant fraction of the energy released is in the form of deka-keV electrons. These electrons \emph{gain} energy through as-yet-not-fully-understood process(es) associated with the reconnection of stressed, current-carrying magnetic fields \citep[see, e.g.,][for a review]{2011SSRv..159..357Z}.  They \emph{lose} energy principally through Coulomb collisions \citep[e.g.,][]{1972SoPh...26..441B,1978ApJ...224..241E}, although ohmic losses associated with driving the beam-neutralizing return current through the finite-resistivity ambient medium
\citep{1977ApJ...218..306K,1980ApJ...235.1055E,1984ApJ...280..448S,1985ApJ...293..584H,1989SoPh..120..343L,1990A&A...234..496V,1995A&A...304..284Z,2005A&A...432.1033Z,2006ApJ...651..553Z,2008A&A...487..337B}, may also be significant.

Early impulsive phase models \citep[e.g.,][]{1971SoPh...18..489B} assumed, largely for simplicity, that the electrons are accelerated out of a ``point source'' at or near the apex of a coronal loop. However, estimates \citep[e.g.,][]{2003ApJ...595L..97H,2011SSRv..159..107H} of the number of accelerated electrons required to produce a strong hard X-ray burst, combined with even generous estimates of the number density of electrons in the acceleration region, show that an acceleration region extending over a substantial portion of the flare volume is required. Additional evidence for an extended acceleration region includes the following:

\begin{itemize}

\item While many observations \citep[e.g.][]{2002ApJ...569..459P,2003ApJ...595L.107E} point to an acceleration site in the corona and the production of hard X-ray footpoints by electrons precipitating into the chromosphere, the appearance of coronal hard X-ray sources \citep[e.g.][]{1995PASJ...47..677M,2003AdSpR..32.2489K,2006ApJ...638.1140J,2006ApJ...648.1239K,2007ApJ...669L..49K,2007A&A...461..315T}, in particular extended nonthermal coronal sources \citep{2004ApJ...603L.117V,2008ApJ...673..576X,2011ApJ...730L..22K,2012A&A...543A..53G,2012ApJ...755...32G,2014ApJ...787...86J} shows that the accelerated electrons are, at least in some events, fully confined to the extended coronal region where the acceleration occurs;

\item {\em RHESSI} \citep{2002SoPh..210....3L} observations reveal that the accelerated electron distribution is nearly isotropic \citep[e.g.,][]{2006ApJ...653L.149K}.  This favors stochastic acceleration mechanisms operating throughout an extended region \citep[e.g.,][]{1992ApJ...398..350H,1994ApJS...90..623M,2008ApJ...687L.111B,2009ApJ...692L..45B,2012SSRv..173..535P,
2012ApJ...754..103B};

\end{itemize}

\citet{2013A&A...551A.135S} have shown that the number of electrons trapped in extended coronal sources significantly exceeds the number consistent with purely collisional transport of these electrons to the chromospheric footpoints.  They thus argue that some form of non-collisional scattering mechanism confines electrons to the coronal part of the loop (and hence to the acceleration region), even in coronal-source-plus-footpoint events. \citet{2014ApJ...787...86J} have noted that observed variation of hard X-ray source length with photon energy is not consistent with a purely collisional transport model but rather with one in which parallel transport proceeds through a process involving both collisions and noncollisional scattering. \citet{2014ApJ...780..176K} further showed, through analysis of similar events, that the mean free path associated with the noncollisional process is of order $10^8$~cm, an order of magnitude or so less than the collisional mean free path.

Local fluctuations in the magnetic field are already well known to be responsible for angular scattering and isotropization of the particle distribution, leading to spatial diffusion along the guiding magnetic field \citep{1966ApJ...146..480J, 1989ApJ...336..243S,2009ASSL..362.....S}.
{\it Cross-field} diffusion of particles in turbulent magnetic fields has also long been considered as a mechanism for cosmic-ray transport \citep{1966ApJ...146..480J,
1969ApJ...155..777J, 2009A&A...507..589S},
for the transport of solar energetic particles \citep{2003ApJ...590L..53M, 2013ApJ...773L..29L},
and for transport of thermal electrons
in coronal loops \citep{2006ApJ...646..615G, 2007A&A...462.1113G, 2010ApJ...719.1912B}.  In this context, \citet{2011ApJ...730L..22K} and \citet{2011A&A...535A..18B} have shown that the width (perpendicular to the guiding magnetic field) of extended coronal hard X-ray sources increases slowly with energy, consistent with transport of energetic electrons across the guiding field lines through collisionless scattering off magnetic inhomogeneities.

In summary, hard X-ray observations of solar flares strongly suggest that electrons are accelerated in an extended region, within which a combination of Coulomb collisions and collisionless pitch-angle scattering operate.  In this paper we explore the implications that the addition of such a non-collisional scattering process has for the transport of electrons of {\it all} energies, both thermal and non-thermal, in the flaring corona.

Soft X-ray observations \citep[e.g.,][ as a recent review]{2011SSRv..159..107H} show that the overall spectrum (and hence the distribution of emitting electrons in the flaring corona) is often near-Maxwellian, with a temperature of a few $\times 10^7$~K. Such fits to flare soft X-ray spectra also provide estimates of the emission measures $EM \equiv \int n^2 \, dV$ of the soft-X-ray-emitting volume; in large flares this can be $\simeq 10^{49}$~cm$^{-3}$ or higher. Since the emitting volume is of order $10^{27}$~cm$^3$, this requires a density of order $10^{11}$~cm$^{-3}$.
At such densities and temperatures, the collisional mean free path is of order $10^8$~cm, significantly less than the characteristic source length $L \sim V^{1/3} \simeq 10^9$~cm.  Thus, unlike in other situations (e.g., in the solar wind), collisional processes are important in determining the ambient conditions in the plasma, which accounts for the very good fit of Maxwellian forms to observed soft X-ray spectra.

In a collisional environment, transport of quantities such as heat and electric charge are principally determined \citep[e.g.][]{1962pfig.book.....S} by local gradients in temperature and density.  The presence of additional non-collisional scattering processes does not change this essentially local {\it nature} of plasma transport phenomena.  However, collisionless scattering processes affect the transport coefficients and hence the values of heat flux and current that arise from prescribed values of the local temperature gradient and large-scale electric field.  In this paper, we evaluate this effect quantitatively, and we also discuss how substantial deviations from the classical \citep{1962pfig.book.....S} values of the thermoelectric transport coefficients can have very significant implications for several areas of importance to the solar flare problem.  Such implications include:

\begin{itemize}
\item{
Suppression of thermal conduction will result in a higher coronal temperature for a prescribed heating rate (e.g., by nonthermal electrons) and so possibly account for very hot sources observed to be confined in the corona \citep[e.g.,][]{1987ApJ...320..904K,1998A&A...334.1112J};}

\item{
A reduction in the thermal conductivity coefficient $\kappa_\parallel$ will lengthen the conductive cooling time $\tau_{\rm cool} \simeq 3nkT/(\kappa_\parallel T^{7/2}/L^2$) from its classical value  \citep{1980sfsl.work..341M} and so offers a possible resolution to the long-standing conundrum of the apparent need for continued energy input to coronal plasma after the impulsive phase.}

\item{
Nonthermal electrons lose energy in driving a beam-neutralizing return current through the finite resistivity of the ambient atmosphere.  For a given beam current density $j_\parallel$, a reduction in the electrical conductivity $\sigma_\parallel$ increases the role of return current losses relative to Coulomb collisions, and this can have significant implications for the spatial distribution of hard X-ray emission and for the energy deposition profile throughout the ambient atmosphere, and hence the hydrodynamic response of the solar atmosphere \citep[e.g.,][]{1984ApJ...279..896N,1989ApJ...341.1067M,2005ApJ...630..573A}.  Confinement of nonthermal electrons through enhanced resistivity may also offer an alternative explanation for loop-top coronal sources \citep{2003AdSpR..32.2489K,2004ApJ...603L.117V,2008ApJ...673..576X,2012A&A...543A..53G,2012ApJ...755...32G,2014ApJ...787...86J}.
And, because the hard X-ray production of a population of accelerated electrons depends inversely on the energy loss rates of the accelerated electrons, the results will change the required total energy content in these accelerated electrons, a quantity of considerable importance in the overall energetics of solar eruptive events \citep{2012ApJ...759...71E}.}
\end{itemize}

In Section~\ref{pitch-angle} we consider the combined effects of collisional and turbulent scattering on the effective mean free path used to compute quantities such as the thermal conductivity $\kappa_\parallel$ and the electrical conductivity $\sigma_\parallel$.  In Section~\ref{modified-spitzer} we study this problem more formally, using a Chapman-Enskog expansion of the kinetic equation for the electron phase-space distribution function $f(z,v,\mu)$, and we derive expressions for the various thermoelectric coefficients that link transport quantities (thermal flux, electrical current density) to the local environment (temperature gradient, electric field).
We do this first for a model of isotropic scattering where the explicit dependence of the diffusion coefficient $D_{\mu\mu}$ on pitch angle $\mu$ is, for simplicity, discarded.
Scattering by magnetostatic fluctuations, which is generally not isotropic, is addressed in Section~\ref{dependent}.
In Section~\ref{application} we discuss the application of these results to several aspects of solar flares, including the heating of coronal plasma by non-thermal electrons and the subsequent cooling of this hot plasma by thermal conduction to the chromosphere (Section~\ref{coronal-heating-cooling}), and the role of return currents in ohmic heating of the corona and in the dynamics of the nonthermal electron population (Section~\ref{return-current}).  In Section~\ref{summary} we summarize the results and present our conclusions.

\section{Effects of Pitch-Angle Scattering on Particle Transport}\label{pitch-angle}

The basis of turbulent scattering theory in plasmas was developed some time ago \citep[see][for a review]{1969npt..book.....S}. Here we adopt the philosophy that turbulence can be thought of as playing a role similar to collisions, resulting in a ``rescaling'' of the transport coefficients with respect to their collisional values. We also focus on elastic angular scattering, which is the predominant effect for scattering of particles by low-frequency turbulence \citep{1966JETP...23..145R}.

\subsection{Modeling the scattering frequency and mean free path}\label{scattering-parameters}

The pitch-angle diffusion coefficient $D_{\mu\mu}$ of charged particles in a magnetized plasma is related to the angular scattering frequency $\nu $ (s$^{-1}$) by

\begin{equation}\label{dmumudef}
D_{\mu\mu}=\nu \, \frac{(1-\mu^2)}{2} \,\,\,,
\end{equation}
where $\mu$ is the cosine of the pitch-angle relative to the guiding magnetic field. In the case where angular scattering is a superposition of two additive processes, collisional and turbulent, the scattering frequency can be written as

\begin{equation}\label{dmumu-total}
\nu (v) = \nu_{C}(v) + \nu_{T}(v) \,\,\, .
\end{equation}
Here the collisional contribution is given by

\begin{equation}\label{dmumu-c}
\nu_{C}(v) = \frac{4\pi n_e \, e^4 \, \ln \Lambda}{m_e^2} \, \frac{1}{v^3} \equiv \frac{v}{\lambda_{\rm C}(v)} \,\,\, ,
\end{equation}
where we have introduced the collisional mean free path

\begin{equation}\label{lambdac-def}
\lambda_{\rm C} (v) = \frac{m_e^2}{4\pi n_e \, e^4 \, \ln \Lambda} \, {v^4}\equiv \lambda_{\rm ei} \, \left( \frac{v}{v_{\rm te}} \right)^{4} \,\,\, .
\end{equation}
Here $e$ and $m_e$ are the electronic charge (esu) and mass (g), respectively, $n_e$ (cm$^{-3}$) is the ambient electron density, $\ln \Lambda$ is the Coulomb logarithm, and the thermal mean free path

\begin{equation}\label{lambda-ei}
\lambda_{\rm ei} = \frac{(2 k_B)^2}{2 \pi e^4 \, \ln \Lambda} \, \frac{T_e^2}{n_e}\simeq \frac{10^{4} \, T_{e}^{2}}{n_{e}} \,\,\, ,
\end{equation}
which, by definition, is the collisional mean free path of electrons with thermal speed $v = v_{\rm te} \equiv \sqrt{2 k_B T_e/m_e}$, where $k_B$ is Boltzmann's constant.

Similarly, the turbulent contribution to the angular scattering frequency may be written

\begin{equation}\label{dmumu-t}
\nu_{T}(v) = \frac{v}{\lambda_{\rm T}(v)} \,\,\, ,
\end{equation}
which involves a (generally velocity-dependent) quantity $\lambda_{\rm T}$, identifiable as the ``turbulent mean free path.'' The physical origin of the turbulence is, for the moment, left unspecified. However, in order to exploit the analogy with collisional scattering we assume a velocity dependence for the turbulent mean free path of the form

\begin{equation}\label{alpha-def}
\lambda_{T}(v)=\lambda_{0} \left( \frac{v}{v_{\rm te}} \right )^{\alpha},
\end{equation}
corresponding to a turbulent scattering frequency

\begin{equation}\label{nut-v}
\nu_{T}(v)=\frac{v} {\lambda_{0}} \, \left ( \frac{v}{v_{\rm te}} \right )^{-\alpha} \,\,\, .
\end{equation}
Using this approach, we can express the total scattering frequency resulting from a combination of collisions and turbulent scattering in the form

\begin{equation}\label{lambda-eff}
\nu (v) = \frac{v}{\lambda(v)} \,\,\, ,
\end{equation}
where the ``effective,'' or simply \emph{the}, mean free path $\lambda (v)$ is given by

\begin{equation}\label{lambda-eff-c-t}
\frac{1}{\lambda(v)} = \frac{1}{\lambda_{\rm C}(v)} + \frac{1}{\lambda_{\rm T}(v)} = \frac{1}{\lambda_{\rm ei}} \left ( \frac{v_{\rm te}}{v} \right )^4 + \frac{1}{\lambda_{\rm T}(v)}
= \frac{1}{\lambda_{\rm ei}} \left ( \frac{v_{\rm te}}{v} \right )^4 + \frac{1}{\lambda_0} \left ( \frac{v_{\rm te}}{v} \right) ^{\alpha} \,\,\, .
\end{equation}
We will find it convenient to introduce the dimensionless ratio

\begin{equation}\label{r}
R = \frac{\lambda_{\rm ei}}{\lambda_{0}} \,\,\, .
\end{equation}
By equations~(\ref{dmumu-c}), (\ref{lambdac-def}), (\ref{nut-v}), and~(\ref{r}), the total scattering frequency can be written in the form

\begin{equation}\label{nu-v}
\nu(v) = \frac{v_{\rm te}}{\lambda_{\rm ei}} \, \frac{1+R (v/v_{\rm te})^{4-\alpha}}{(v/v_{\rm te})^3} \,\,\, ,
\end{equation}
with the corresponding expression

\begin{equation}\label{lambda-v}
\lambda(v) = \frac{\lambda_{\rm ei} \, (v/v_{\rm te})^4}{1+R (v/v_{\rm te})^{4-\alpha}}
\end{equation}
for the mean free path.

The parameter $R$ is the ratio of two length scales, collisional and turbulent, and here plays the role of a transport reduction factor. In the presence of turbulence, the scattering frequency for thermal electrons with $v\sim v_{\rm te}$ is $\nu=\nu_{\rm ei}(1+R)$, an increase by a factor $(1+R)$ relative to the collisional value $\nu_{\rm ei}=v_{\rm te}/\lambda_{\rm ei}$. When $R$ is large, this increased scattering frequency corresponds to a decrease of the the mean free path by a factor $\simeq$$R$.

Pitch-angle scattering due to turbulence is in general \emph{not} isotropic; the scattering frequency may also depend on pitch angle, say as

\begin{equation}\label{nu_t_anisotropic}
\nu_{T}(v,\mu) = \frac{v} {\lambda_{0}} \, \left ( \frac{v}{v_{\rm te}} \right )^{-\alpha}|\mu|^{\beta} \,\,\, .
\end{equation}
In such cases, the mean free path is related to the scattering frequency by the relation

\begin{equation}\label{oo}
\lambda_{T} (v,\mu) =
\frac{3v}{4}\int _{-1}^{+1} \, \frac{(1-\mu^{2})}{\nu(v,\mu)} \, d\mu \,\,\, .
\end{equation}
We caution that in the limit of very large $R$, one cannot compute the mean free path simply by first taking this limit in Equation~(\ref{nu-v}) and then substituting this into Equation~(\ref{oo}). For example, it is easily checked that, for pure turbulent scattering ($\nu=\nu_{T}$) and certain values of $\beta$ (e.g., $\beta=1$), the expression~(\ref{oo}) for the purely turbulent mean free path diverges. Physically, this is because for such values of $\beta$ turbulent scattering alone is incapable of scattering particles through the $90^\circ$ ($\mu=0$) ``barrier.'' It is therefore essential, in general, to include collisional effects even in the limit of large $R=\lambda_{\rm ei}/\lambda_{0}$, so that the mean free path given by Equation~(\ref{oo}) remains finite. The corresponding scattering frequency is given (cf. Equation~(\ref{nu-v})) by

\begin{equation}\label{nu-v-3}
\nu(v,\mu) = \frac{v_{\rm te}}{\lambda_{\rm ei}} \, \frac{1+R \, |\mu|^{\beta} \, (v/v_{\rm te})^{4-\alpha}}{(v/v_{\rm te})^3} \,\,\, .
\end{equation}

The existence of a turbulent spectrum of magnetic fluctuations transverse to the guiding magnetic field is well-documented \citep[see, e.g.,][for a review]{2009ASSL..362.....S}. Such fluctuations can be considered as quasi-static provided the characteristic particle velocity (a few $v_{\rm te}$ for the electrons that carry the bulk of the thermal flux) is larger than $\lambda _{B}/\tau_{B}$, where $\lambda_{B}$ and $\tau_{B}$ are respectively the correlation length and time associated with the fluctuations. Typically, the turbulent mean free path parameter $\lambda_{0}$ is proportional to the inverse square of the fractional level of the magnetic fluctuations, i.e.,

\begin{equation}\label{lambda0-lambdaB}
\lambda_{0} \simeq \lambda_{B} \, \left ( \frac{\delta B_{\perp}}{B_{0}} \right )^{-2},
\end{equation}
so that the value of the ratio $R$ is

\begin{equation}
R = \frac{\lambda_{\rm ei}}{\lambda_{0}} \simeq \left ( \frac{\lambda_{\rm ei}}{\lambda_B} \right ) \,
\left ( \frac{\delta B_\perp}{B_0} \right )^2 = \frac{ 10^4 \, T_e^2 \, (\delta B_\perp/B_0)^2}{n_e \, \lambda_B} \,\,\, .
\end{equation}

\subsection{Effect on transport coefficients}

We now consider in a semi-quantitative way the impact of adding turbulent scattering on various transport coefficients.
The ``collisionality'' of a plasma is inversely proportional to the ratio of the collisional mean free path $\lambda_C$ to the temperature gradient length scale $L_T$; low values the collisional Knudsen number ${\rm Kn}_S \equiv \lambda_C/L_T$ imply a high degree of collisionality and vice versa.  (Here the subscript $S$ stands for \citet{1962pfig.book.....S}.) The collisionality of the solar electron population ranges from quite high (${\rm Kn}_{S}\sim 10^{-2}$) close to the Sun to very low (${\rm Kn}_{S}\sim 1$) in the solar wind at 1~AU. For coronal loops, $L_{\rm T}$ is of the order of the loop length $L$, therefore the collisional Knudsen number  ${\rm Kn}_{S} \simeq 10^{4}T_{e}^{2}/n_{e}L$.  Taking $n_{e} = 10^{10}$~ cm$^{-3}$ and $L=10^{9}$~cm, the collisional Kundsen number ${\rm Kn}_{S}=10^{-15}T_{e}^{2}$, meaning that the range of temperatures $T_{e}=10^{6}-10^{7}$~K corresponds to collisional Knudsen numbers ${\rm Kn}_{S}=10^{-3}-10^{-1}$.

Now let us consider the heat flux density carried by electrons along the field line, which for free-streaming electrons may be straightforwardly written as

\begin{equation}\label{qparallel}
q_{\parallel} = n_{e} m_e v_{\rm te}^3 = n_e m_e \left ( \frac{2 k_B T_e}{m_e} \right )^{3/2} \,\,\, .
\end{equation}
However, we know from Fick's law that when the collisional Knudsen number $\lambda/L_T$ is small, the thermal flux is driven by the local temperature gradient.  Hence we write $T^{3/2}_{e} = - \lambda_C \, T_{e}^{1/2}dT_e/dz$, resulting in an approximate expression for the parallel heat flux in a collisional environment:

\begin{equation}\label{kappa-def-coll}
q_{\parallel} = -\frac{2 n_e \, k_B \, (2 k_B T_e)^{1/2}}{m_e^{1/2}} \, \lambda_C \, \frac{dT_e}{dz} \equiv - \, \kappa_{\parallel,S} \, \frac{dT_e}{dz} \,\,\, .
\end{equation}
Adding collisionless scattering effects gives

\begin{equation}\label{kappa-def}
q_{\parallel} = -\frac{2 n_e \, k_B \, (2 k_B T_e)^{1/2}}{m_e^{1/2}} \, \lambda \, \frac{dT_e}{dz} \equiv - \, \kappa_\parallel \, \frac{dT_e}{dz} \,\,\, ,
\end{equation}
where

\begin{equation}
\kappa_{\parallel}=\frac{2 n_e \, k_B \, (2 k_B T_e)^{1/2}}{m_e^{1/2}} \, \lambda
\end{equation}
is the thermal conductivity, which, through the form of $\lambda$ (Equation~(\ref{lambda-v})), includes the effects of both collisional and non-collisional scattering.

When the turbulent transport reduction factor $R = \lambda_{\rm ei}/\lambda_{0}$ is small, the mean free path takes on its collisional value $\lambda \simeq \lambda_{\rm ei}$ and the parallel heat conductivity correspondingly assumes the standard \citep[e.g.,][]{1962pfig.book.....S} collisional form $\kappa_{\parallel S} = 2 n_e k_B \, (2 k_B T_e)^{1/2} \, \lambda_{\rm ei}/m_e^{1/2} = k_B (2 k_B T_e)^{5/2}/(\pi m_e^{1/2} e^4 \ln \Lambda)$.  (Note that this is density independent because $\lambda_{ei} \propto T_e^2/n_e$.) On the other hand, when the turbulent reduction factor $R$ is large, the heat conductivity becomes significantly suppressed relative to the collisional value and obeys the (generally density dependent) scaling $\kappa_\parallel = 2 n_e k_B (2 k_B T_e)^{1/2} \lambda_{0}/m_e^{1/2} = R^{-1} \kappa_{\parallel S}$.

Next we consider the {\it electrical} conductivity in a turbulent plasma.  As usual, this can be obtained by writing the balance between the electric force and the friction force acting on an electron:

\begin{equation}\label{eqn-motion}
-eE_\parallel - \nu \, m_e v_\parallel = 0 \,\,\, ,
\end{equation}
where $E_\parallel$ (statvolt~cm$^{-1}$) is the component of the electric field in the direction of the background guiding magnetic field. From this we find the parallel current density (defined by $j_\parallel = - e n_e v_\parallel$) to be $j_\parallel = (n_e e^2/\nu \, m_e) \, E_\parallel = (n_e e^2 \lambda/m_e v_{\rm te}) \, E_\parallel$, leading to the Ohm's law

\begin{equation}\label{sigma-def}
j_\parallel = \frac{n_{e} e^2 \lambda}{m^{1/2}_e (2 k_B T_e)^{1/2}} \, E_\parallel \equiv \sigma_\parallel \, E_\parallel \,\,\, ,
\end{equation}
where the electrical conductivity

\begin{equation}
\sigma_\parallel=\frac{n_{e} e^2 \lambda}{m^{1/2}_e (2 k_B T_e)^{1/2}} \,\,\, .
\end{equation}
Again, when $R\ll 1$, the mean free path takes on its collisional value and the parallel electric conductivity obeys the collisional scaling $\sigma_{\parallel S}=n_e e^2 \lambda_{\rm ei}/m_e v_{\rm Te} = (2 k_B T_e)^{3/2}/(2 \pi m_e^{1/2} e^2 \ln \Lambda)$. On the other hand, when the turbulent reduction factor $R\gg 1$, the electric conductivity becomes significantly suppressed relative to the collisional value and obeys the scaling $\sigma_\parallel = n_e e^2 \lambda_{0}/m_e v_{\rm te} = R^{-1} \, \sigma_{\parallel S}$.

We may summarize the above discussion into one simple and rather obvious expression, valid for $v \simeq v_{\rm te}$,

\begin{equation}\label{dparallel-ratio}
\frac{\nu_{\rm ei}}{\nu}\sim \frac{\lambda}{\lambda_{\rm ei}}\sim
\frac{{\rm Kn}}{{\rm Kn}_{S}}\sim \frac{D_{\parallel}}{D_{\parallel S}}\sim\frac{\kappa_{\parallel}}{\kappa_{\parallel S}}\sim\frac{\sigma_{\parallel}}{\sigma_{\parallel S}}\sim \frac{1}{1+R} \,\,\, .
\end{equation}
An enhanced rate of angular scattering yields reduced conductivities compared with the collisional (Spitzer) values.

\section{The modified Spitzer problem}\label{modified-spitzer}

In this section we will improve upon the dimensional estimates above by using a a more rigorous Chapman-Enskog expansion of the electron kinetic equation, i.e., by solving a standard \citep{1962pfig.book.....S} collisional transport problem that includes an additional non-collisional source of pitch-angle scattering. The model will involve two important non-dimensional parameters: the Knudsen number ${\rm Kn}$, which measures the degree of collisionality vs. free-streaming and is assumed to be smaller than unity, and the turbulent reduction factor $R$ (as defined above), which measures the additional role of collisionless scattering. Note that these non-dimensional numbers are independent; thus the large-$R$ asymptotic solutions should not be confused with the collisionless limit ${\rm Kn} \rightarrow \infty$: collisions are always essential (even when $R\gg 1$) to drive the electrons toward the Maxwellian distribution and to keep the overall mean free path finite.

\subsection{Isotropic scattering}

We first consider the impact of an additional source of \emph{isotropic} scattering on the transport coefficients. (This will set up the framework for later addressing the problem of angular scattering by a spectrum of transverse magnetostatic fluctuations, which is generally not isotropic.) Let us consider, then, the one-dimensional kinetic equation for a gyrotropic ($\partial/\partial \phi = 0$) electron distribution function $f(z,\theta, v, t)$ under the action of an electric field $E_\parallel$ parallel to the ambient magnetic field ${\bf B}$,

\begin{equation}\label{kinetic}
\frac{\partial f}{\partial t} + v_\parallel \, \mathbf{b}.\nabla f - \frac{eE_\parallel}{m_e} \, \mathbf{b}.\nabla_{\mathbf{v}}f = St^{v}(f) + St^{\theta}(f) \,\,\, ,
\end{equation}
where $z$ (cm) is the position of the particle gyrocenter along the magnetic field with direction ${\mathbf b} = {\mathbf B_0}/B_0$, $\theta$ is the pitch angle ($\cos \theta = \mathbf {v}.\mathbf{B}_0/vB_0 = v_\parallel/v$) and $v = \sqrt{v_\parallel^2 + v_\perp^2}$ is the electron speed. Transforming to the variables $(z,\mu,v,t)$, with $\mu= \cos \theta$ being the pitch-angle cosine, this Fokker-Planck equation may be rewritten as

\begin{equation}\label{fokker}
\frac{\partial f}{\partial t} + \mu \, v \, \frac{\partial f}{\partial z} - \frac{e E_\parallel}{m_e} \, \mu\, \frac{\partial f}{\partial v} - \frac{e E_\parallel}{m_e} \, \frac{(1-\mu^2)}{v}
\, \frac{\partial f}{\partial \mu} = St^{v}(f) + St^{\mu}(f) \,\,\, .
\end{equation}
The velocity scattering operator

\begin{equation}\label{stv}
St^{v}(f) = \frac{1}{v^{2}} \, \frac{\partial}{\partial v} \left [ v^{2} D(v) \left ( \frac{\partial f}{\partial v} + \frac{m_e}{k_B T_e} \, fv \right ) \right ]
\end{equation}
describes collisional diffusion in velocity space and collisional drag. The diffusion coefficient in velocity space is given by

\begin{equation}\label{diff-v}
D(v) = \frac{4 \pi e^{4} \ln \Lambda \, n_e k_B T_e}{m_e^3} \, \frac{1}{v^3} \,\,\, ,
\end{equation}
whereas the pitch-angle scattering operator takes the form

\begin{equation}\label{stmu}
St^{\mu}(f) =  \frac{\partial}{\partial \mu} \left [ D_{\mu\mu}\, \frac{\partial f}{\partial \mu} \right ] =\frac{\partial}{\partial \mu} \left [ \frac{\nu (v)}{2} \, (1-\mu^{2}) \, \frac{\partial f}{\partial \mu} \right ] \,\,\, .
\end{equation}
While the velocity-space scattering operator~(\ref{stv}) is responsible for kinetic energy change, the pitch-angle scattering operator~(\ref{stmu}) is responsible for momentum change at constant kinetic energy. We recall that the scattering frequency $\nu(v)$, and hence (Equation~(\ref{dmumudef})) the pitch-angle diffusion coefficient $D_{\mu\mu}$, are sums of two components -- collisional and turbulent pitch-angle diffusion; thus $D_{\mu\mu} = D^C_{\mu\mu} + D^T_{\mu\mu}$.

The right hand side of Equation~(\ref{fokker}) vanishes identically for a Maxwellian distribution, i.e.,

\begin{equation}\label{st-maxwellian}
St^{v}(f_{0}) + St^{\mu}(f_{0})  =0 \,\,\, ,
\end{equation}
where

\begin{equation}\label{maxwellian}
f_{0}(v) = n_e \left ( \frac{m_e}{2\pi k_B T_e} \right )^{3/2} \, e^{-m_e v^2/2k_B T_e} \,\,\, .
\end{equation}
However, if $T$ and/or $n$ are not uniform -- $T_e=T_e(z)$ and/or $n_e=n_e(z)$ -- then the (local) Maxwellian does {\it not} cancel the spatial transport term on the left hand side of the Fokker-Planck equation~(\ref{fokker}). This deviation from the homogeneous equilibrium state is responsible for the spatial
flux of particles in the plasma and of what they carry (e.g., kinetic energy, electric charge). In this situation, we use a standard Chapman-Enskog expansion of the electron kinetic equation and write the distribution function as the sum of a zeroth order isotropic Maxwellian distribution plus a small flux-carrying anisotropic correction:

\begin{equation}\label{perturb}
f = f_0(z,v) + \epsilon f_1(z,\mu,v) \,\,\, ,
\end{equation}
where the expansion parameter $\epsilon$ is of the order of the (small) Knudsen number in the plasma, i.e., $\epsilon \sim$~Kn. All operators being linear, we obtain at order $\epsilon$ an equation for $f_{1}$:

\begin{equation}\label{first-order}
St^{v}(f_{1}) + \frac{\partial}{\partial \mu} \left [ D_{\mu\mu}(v) \, \, \frac{\partial f_1}{\partial \mu} \right ] =
\mu \, v \, \frac{\partial f_0(z, v)}{\partial z} - \mu \, \left ( \frac{e E_\parallel}{m_e} \right ) \,
\frac{\partial f_0(z, v)}{\partial v} \,\,\, ,
\end{equation}
where $f_0$ is the Maxwellian distribution~(\ref{maxwellian}). By solving this equation for the lead anisotropic component $f_1$, we can obtain expressions for both the heat flux and the electric current. This constitutes a standard \citet{1962pfig.book.....S} problem which becomes easily tractable in the Lorentz plasma approximation\footnote{This approximation is, for the Coulomb interaction, equivalent to the Lorentz gas approximation which has immobile heavy hard spheres as scattering agents, as in the Drude-Lorentz model of electric conductivity, hence the name ``Lorentz plasma'' \citep[e.g.,][]{1964PhFl....7..407K},
see also \citep{1966JETP...23..145R} for the analogy between low-frequency electrostatic turbulence and the Lorentz gaz}, i.e., if one assumes that the scattering is angular only, i.e., $St^v(f_1)=0$.  For such a case, Equation~(\ref{first-order}) can be immediately integrated once over $\mu$ to give

\begin{equation}\label{df1dmu-c}
\frac{\partial f_1}{\partial \mu} = \frac{\mu^2}{\nu \, (1-\mu ^2)} \, \left [ v \, \frac{\partial f_0(z, v)}{\partial z}-
\frac{e E_\parallel}{m_e} \, \frac{\partial f_0(z, v)}{\partial v} \right ] + \frac{C}{\nu \, (1-\mu ^2)} \,\,\, .
\end{equation}
The choice of the integration constant

$$C = -  \left [ v \, \frac{\partial f_{0}(z, v)}{\partial z} - \frac{e E_\parallel}{m_e} \, \frac{\partial f_0(z, v)}{\partial v} \right ]$$
yields

\begin{equation}\label{df1dmu}
\frac{\partial f_1}{\partial \mu} = -\frac{1}{\nu } \left [ v \, \frac{\partial f_0(z, v)}{\partial z}- \frac{e E_\parallel}{m_e} \, \frac{\partial f_0(z, v)}{\partial v} \right ] \,\,\, ,
\end{equation}
which ensures the regularity of $f_1$ at $\mu=\pm 1$.

Since we are, for now, adopting an isotropic scattering model in which the scattering frequency $\nu$ is independent of the pitch angle cosine $\mu$, one can integrate Equation~(\ref{df1dmu}) once more to obtain

\begin{equation}\label{f1-general}
f_1(z, v, \mu) = - \frac{\mu}{\nu} \left [ v \, \frac{\partial f_0(z, v)}{\partial z}- \frac{e E_\parallel}{m_e} \, \frac{\partial f_0(z, v)}{\partial v} \right ] \,\,\, .
\end{equation}
Evaluating the pertinent space and velocity derivatives of $f_0$, and assuming a constant pressure along $z$: $n_e(z) k_B T_e(z) = P_e = {\rm constant}$, we obtain

\begin{equation}\label{f1-basic}
f_1 (z, v, \mu) = - \frac{\mu  v}{\nu} \left [ \left ( \frac{m_ev^2}{2 k_B T_e} - \frac{5}{2} \right )
\frac{1}{T_e} \, \frac{d T_e}{dz} + \frac{eE_\parallel}{k_B T_e} \right ] \, f_0 \,\,\, ,
\end{equation}
with

\begin{equation}\label{dmumu-v}
\nu (v) = \frac{v}{\lambda_{\rm c}(v)} + \frac{v}{\lambda_{\rm T}(v)} = \frac{v_{\rm te}}{\lambda_{\rm ei}} \, \frac{1+R(v/v_{\rm te})^{4-\alpha}}{(v/v_{\rm te})^3} \,\,\, .
\end{equation}
It is convenient to introduce the normalization

\begin{equation}\label{x-def}
x = \frac{v}{v_{\rm te}} \,\,\, ,
\end{equation}
so that the zero-order Maxwellian distribution (Equation~(\ref{maxwellian})) at a given location (i.e., given values of $T_e$ and $v_{\rm te}$) can be written in the form

\begin{equation}\label{f0}
f_0(x) = \pi^{-3/2} n_{e} \, v_{\rm te}^{-3} \, e^{-x^{2}} \,\,\, .
\end{equation}
Further, Equation~(\ref{dmumu-v}) becomes

\begin{equation}\label{dmumux3}
\nu = \frac{v_{\rm te}}{\lambda_{\rm ei}} \, \frac{Rx^{4-\alpha}+1}{x^{3}} \,\,\, ,
\end{equation}
and thus the expression~(\ref{f1-basic}) for $f_1$ becomes

\begin{equation}\label{f1-x-mu}
f_1 = - \mu \lambda_{\rm ei} \, \frac{x^4}{Rx^{4-\alpha}+1} \, \left [ \left ( x^2-\frac{5}{2} \right )
\frac{1}{T_e} \, \frac{dT_e}{dz} + \frac{eE_\parallel}{k_B T_e} \right ] \, f_0 \,\,\, .
\end{equation}
We can now compute the heat flux from

\begin{equation}\label{heat-flux-def}
q_\parallel(z) = 2 \pi \, \int_0^\infty dv \, v^2 \int_{-1}^1 d\mu \, \mu \, \left ( \frac{m_e v^2}{2} \right ) \, v \, f_1 (z, v, \mu) \,\,\, ,
\end{equation}
which yields

\begin{eqnarray}\label{qparallel-mu-v-f1}
q_\parallel(z)
&=& 2\pi k_B T_e \, v_{\rm te}^4\int_0^\infty dx \, x^5 \int_{-1}^1 d\mu \, \mu \, f_1(z,x,\mu)
\cr
&=& - \frac{4}{3 \sqrt{\pi}} \, n_{e} \, k_B T_e \, v_{\rm te} \, \lambda_{\rm ei}
\int_0^\infty dx \, \frac{x^9}{Rx^{4-\alpha}+1} \, \left [ \left ( x^2-\frac{5}{2} \right )
\frac{1}{T_e} \, \frac{dT_e}{dz}+\frac{eE_\parallel}{k_B T_e} \right ] \, e^{-x^2} \,\,\, .
\end{eqnarray}
Writing this in the form

\begin{equation}\label{qparallel-onsager-2}
q_\parallel = - \kappa_\parallel \, \frac{dT_e}{dz} - \alpha_\parallel \, E_\parallel \,\,\, ,
\end{equation}
gives the thermoelectric coefficients

\begin{equation}\label{kappa-parallel}
\kappa_{\parallel} = \frac{4}{3\sqrt{\pi}} \,  n_e \, k_B \, v_{\rm te} \, \lambda_{\rm ei} \int_0^\infty \frac{x^9}{Rx^{4-\alpha}+1} \, \left ( x^2-\frac{5}{2} \right ) \, e^{-x^2} \, dx
\end{equation}
and

\begin{equation}\label{alpha-parallel}
\alpha_{\parallel} = \frac{4}{3\sqrt{\pi}} \,  n_e \, e \, v_{\rm te} \, \lambda_{\rm ei} \int_0^\infty \frac{x^9}{Rx^{4-\alpha}+1} \, e^{-x^2} \, dx \,\,\, ,
\end{equation}
respectively.

Similarly, substituting for $f_1$ from Equation~(\ref{f1-basic}) in the expression for the parallel current density

\begin{equation}\label{current-density-def}
j_\parallel(z) = - 2 \pi \, e \int_0^\infty dv \, v^2 \int_{-1}^1 d\mu \, \mu \, v \, f_1(z, v, \mu)
\end{equation}
gives

\begin{eqnarray}\label{jparallel-mu-v-f1}
j_\parallel(z)
&=& - 2 \pi e \, v_{\rm te}^4 \int dx \, x^3 \int_{-1}^1 d\mu \, \mu \, f_1(z,x,\mu)
\cr
&=& \frac{4}{3 \sqrt{\pi}} \,  n_e \, e \, v_{\rm te} \, \lambda_{\rm ei} \, \int_0^\infty dx \, \frac{x^7}{Rx^{4-\alpha}+1} \, \left [ \left ( x^2-\frac{5}{2} \right )
\frac{1}{T_e} \, \frac{dT_e}{dz} + \frac{eE_\parallel}{k_B T_e} \right ] \, e^{-x^2} \,\,\, ,
\end{eqnarray}
and writing this in the form

\begin{equation}\label{jparallel-onsager-2}
j_\parallel = \beta_\parallel \, \frac{dT_e}{dz} + \sigma_\parallel \, E_\parallel
\end{equation}
gives the identifications

\begin{equation}\label{beta-parallel}
\beta_{\parallel} = \frac{4}{3\sqrt{\pi}} \,  \frac{n_e \, e \, v_{\rm te} \, \lambda_{\rm ei}}{T_e} \int_{0}^{\infty} \frac{x^7}{Rx^{4-\alpha}+1} \, \left ( x^2-\frac{5}{2} \right ) \, e^{-x^2} \, dx
\end{equation}
and

\begin{equation}\label{sigma-parallel}
\sigma_{\parallel}  = \frac{4}{3\sqrt{\pi}} \, \frac{n_e \, e^2 \, v_{\rm te} \, \lambda_{\rm ei}}{k_B T_e} \int_0^\infty \frac{x^7}{Rx^{4-\alpha}+1} \, e^{-x^2} \, dx \,\,\, .
\end{equation}
Note that enforcing the current neutrality condition $j_\parallel=0$ in Equation~(\ref{jparallel-onsager-2}) shows that $E_\parallel = (-\beta_\parallel/\sigma_\parallel) \, dT_e/dz$.  Substituting this condition into Equation~(\ref{qparallel-onsager-2}) implies that the thermal conductivity coefficient in a current
 neutralized environment is given by $\kappa^* = \kappa_\parallel - \alpha_\parallel \beta_\parallel/\sigma_\parallel$.

We now provide explicit expressions for these coefficients in the $R \ll 1$ (collision-dominated) and $R \gg 1$ (turbulence-dominated) limiting regimes.

\subsection{Limit of small R}
As expected, in the limit $R\ll 1$ we can replace the term $(Rx^{4-\alpha}+1)$ in the denominators of the integrands in expressions~(\ref{kappa-parallel}), (\ref{alpha-parallel}), (\ref{beta-parallel}), and (\ref{sigma-parallel}) with unity, thus recovering the collisional \citep{1962pfig.book.....S} values. Noting that the integral

\begin{equation}
\int_0^\infty dx \, x^n \, e^{-x^2} = \frac{1}{2} \, \Gamma \left ( \frac{n + 1}{2} \right ) = \frac{1}{2} \, \left (\frac{n-1}{2} \right ) !
\end{equation}
gives the purely collisional results

\begin{equation}\label{kappa-sn}
\kappa_{\parallel} = \frac{2\Gamma(6)- 5\Gamma(5)}{3 \sqrt{\pi}} \, n_e k_B \, v_{\rm te} \, \lambda_{\rm ei} = \frac{40}{\sqrt{\pi}} \, n_e k_B \, v_{\rm te} \, \lambda_{\rm ei} \,\,\, ;
\end{equation}

\begin{equation}\label{alpha-sn}
\alpha_{\parallel} = \frac{2 \, \Gamma(5)}{3 \sqrt{\pi}} \, n_e \, e \, v_{\rm te} \, \lambda_{\rm ei} = \frac{16}{\sqrt{\pi}} \, n_e \, e \, v_{\rm te} \, \lambda_{\rm ei} \,\,\, ;
\end{equation}

\begin{equation}\label{beta-sn}
\beta_{\parallel}=\frac{2\Gamma (5)-5\Gamma(4)}{3 \sqrt{\pi}} \, \frac{n_e \, e \, v_{\rm te} \, \lambda_{\rm ei}}{T_e}=\frac{6}{\sqrt{\pi}} \, \frac{n_e \, e \, v_{\rm te} \, \lambda_{\rm ei}} {T_e} \,\,\, ;
\end{equation}
and

\begin{equation}\label{sigma-sn}
\sigma_{\parallel} =\frac{2 \, \Gamma(4)}{3 \sqrt {\pi}} \frac{n_e e^2 \, v_{\rm te} \, \lambda_{\rm ei}}{k_B T_e} = \frac{4}{\sqrt {\pi}} \frac{n_e e^2 \, v_{\rm te} \, \lambda_{\rm ei}}{k_B T_e} \,\,\, .
\end{equation}
Finally, in the zero-current regime, the effective thermal conductivity coefficient is

\begin{equation}\label{kappa-star-sn}
\kappa^{*} \equiv \kappa_\parallel - \frac{\alpha_\parallel \beta_\parallel}{\sigma_\parallel} = \frac{16}{\sqrt{\pi}} \, n_e k_B \, v_{\rm te} \, \lambda_{\rm ei} \,\,\, .
\end{equation}

\subsection{Limit of large R}

On the other hand, when $R\gg 1$ we replace the term $(Rx^{4-\alpha}+1)$ in the denominators of the integrands in expressions~(\ref{kappa-parallel}), (\ref{alpha-parallel}), (\ref{beta-parallel}), and (\ref{sigma-parallel}) with $Rx^{4-\alpha}$.  This gives the following expressions:

\begin{equation}\label{kappa-sn-rgg1}
\kappa_{\parallel} = \frac{2\Gamma(\frac{8+\alpha}{2})-5\Gamma(\frac{6+\alpha}{2})}{3 \sqrt{\pi}} \,\, \frac{n_e k_B \, v_{\rm te} \, \lambda_{\rm ei}}{R}\,\,\, ,
\end{equation}

\begin{equation}\label{alpha-sn-rgg1}
\alpha_{\parallel} = \frac{2 \, \Gamma(\frac{6+\alpha}{2})}{3 \sqrt{\pi} \, R} \, \, n_e \, e \, v_{\rm te} \, \lambda_{\rm ei} \,\,\, ,
\end{equation}

\begin{equation}\label{beta-sn-rgg1}
\beta_{\parallel} = \frac{2\Gamma(\frac{6+\alpha}{2})-5\Gamma(\frac{4+\alpha}{2})}{3 \sqrt{\pi} \, R} \,\, \frac{n_e \, e \, v_{\rm te} \, \lambda_{\rm ei}}{T_e}\,\,\, ;
\end{equation}
and

\begin{equation}\label{sigma-sn-rgg1}
\sigma_{\parallel} = \frac{2 \, \Gamma(\frac{4+\alpha}{2})}{3 \sqrt {\pi} \, R} \,\, \frac{n_e e^2 \, v_{\rm te} \, \lambda_{\rm ei}}{k_B T_e}\,\,\, ;
\end{equation}
The case $\alpha=0$ is of special interest as it corresponds (Equation~(\ref{alpha-def})) to a velocity-independent turbulent mean free path $\lambda_T$. In this case we have

\begin{equation}\label{kappa-sn-rgg1alpha0}
\kappa_{\parallel}=\frac{2}{3\sqrt{\pi} \, R} \, n_e k_B \, v_{\rm te} \, \lambda_{\rm ei} \,\,\, ,
\end{equation}

\begin{equation}
\alpha_{\parallel}= \frac{4}{3 \sqrt{\pi} \, R} \, n_e \, e \, v_{\rm te} \, \lambda_{\rm ei}  \,\,\, ,
\end{equation}

\begin{equation}\label{beta-sn-rgg1alpha0}
\beta_{\parallel} = - \frac{1}{3 \sqrt{\pi} \, R} \, \frac{n_e \, e \, v_{\rm te} \, \lambda_{\rm ei}}{T_e} \,\,\, ;
\end{equation}
and

\begin{equation}\label{sigma-sn-rgg1alpha0}
\sigma_{\parallel} = \frac{2}{3 \sqrt {\pi} \, R} \, \frac{n_e e^2 \, v_{\rm te} \, \lambda_{\rm ei}}{k_B T_e} \,\,\, .
\end{equation}
In a zero-current scenario, the effective thermal conductivity coefficient is

\begin{equation}\label{kappa-star-sn-rgg1alpha0}
\kappa^{*} \equiv \kappa_\parallel - \frac{\alpha_\parallel \beta_\parallel}{\sigma_\parallel} = \frac{4}{3 \sqrt{\pi} \, R} \, n_e k_B \, v_{\rm te} \, \lambda_{\rm ei} \,\,\, .
\end{equation}
As expected from the simple dimensional arguments of the previous Section, in the turbulence-dominated limit of large $R$ the transport coefficients are suppressed by a factor of $R$ with respect to the Spitzer values.

\begin{figure}[pht]
\centering
\includegraphics[width=0.56\linewidth]{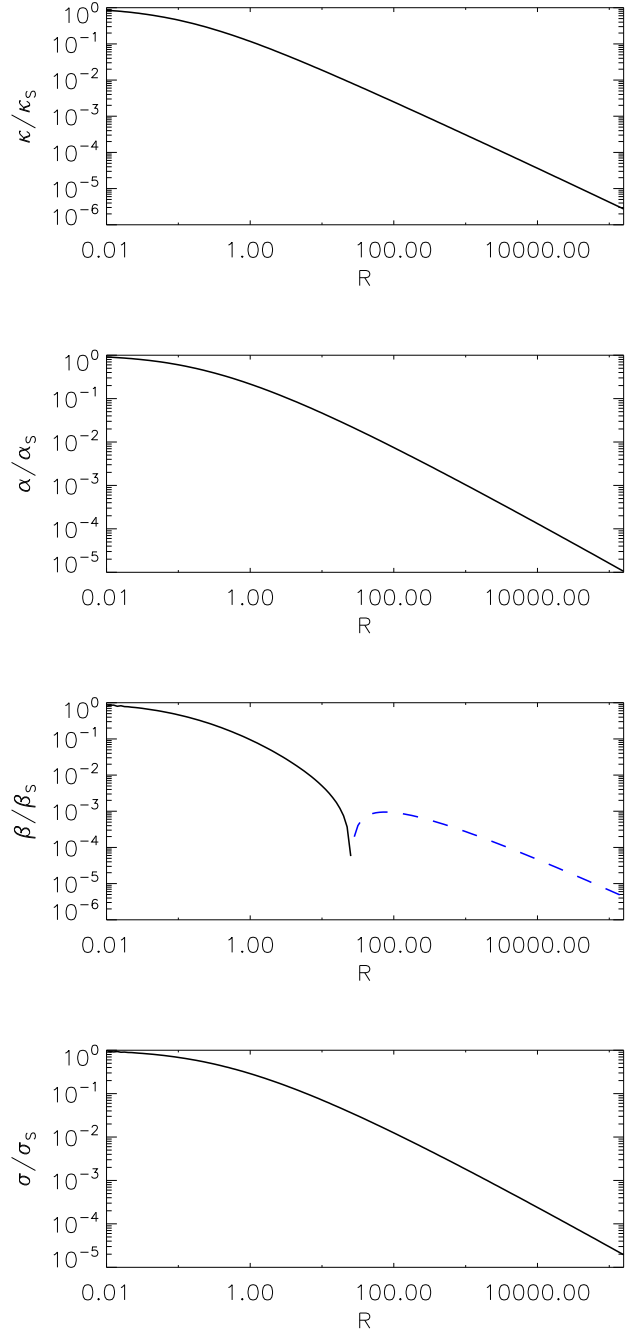}
\caption{Transport coefficients values, relative to their collisional (Spitzer; $R=0$) limit as a function of $R$ for the case $\alpha = 0$ (turbulent mean free path $\lambda_T$ independent of velocity). The blue dashed line indicates negative values, which appear only for $\beta_\parallel$ in the high-$R$ limit; see explanation in the text.}
\label{fig:coefficients}
\end{figure}

Figure~\ref{fig:coefficients} shows the full dependence on $R$ of all four thermoelectric coefficients for the case $\alpha=0$.  We note from this Figure that in the turbulence-dominated high-$R$ limit, one cross-transport coefficient ($\beta_\parallel$) has not only changed in magnitude, but also in {\it sign}, compared to its collisional value (compare Equations~(\ref{beta-sn}) and~(\ref{beta-sn-rgg1alpha0})). Physically, this sign reversal is a consequence of the form of the Maxwellian distribution.  In a uniform pressure environment,

\begin{equation}\label{f0-T-dependence}
f_0(z,v) \sim \frac{n_e(z)}{[T_e(z)]^{3/2}} e^{-mv^2/2 k_B T_e(z)} = \left ( \frac {P_e}{k_B} \right ) \,
[T_e(z)]^{-5/2}  \, e^{-mv^2/2 k_B T_e(z)} \,\,\, ,
\end{equation}
from which we see that there are two contributions to the temperature gradient $dT_e/dz$, one involving the derivative of the normalization coefficient $T_e^{-5/2}$ and the other the derivative of the width of the Maxwellian velocity profile.  The relative signs of these terms depend on the value of the speed $v$: increasing the temperature flattens the Gaussian velocity profile, leading to a general increase in $f_0$ at a high speeds but a general decrease at low speeds.  As a consequence, the spatial derivative

\begin{equation}\label{df0bdz}
\frac{\partial f_0(z,v)}{\partial z} = \frac{\partial f_0(z,v)}{\partial T_e} \, \frac{dT_e}{dz} \,\,\, ,
\end{equation}
that appears in the expression~(\ref{f1-general}) for the lead anisotropic component $f_1(z,v,\mu)$ changes sign at some value of $v$.  The overall sign of a particular transport coefficient (e.g., $\kappa_\parallel$, $\beta_\parallel$) depends on the relative contributions of the particular moments of $f_1(v,z,\mu)$ that appear\footnote{Note that only one term arises in evaluating the velocity-space derivative $\partial f_0(z,v)/\partial v$ that prefixes the electric field $E_\parallel$ term in the expression for $f_1(z,v,\mu)$. Thus only one velocity-moment term appears in each of the expressions for $\alpha_\parallel$ and $\sigma_\parallel$, and the signs of these two thermoelectric coefficients are therefore fixed.} in the expression for that coefficient.  Higher-order moments of the velocity distribution are dominated by larger values of $v$, so that an increase in temperature generally leads to an increase in that moment, while lower-order moments place a greater weight on lower values of $v$, leading to a reduced increase (or even a decrease) in that moment.  And, since (1) the moments of the velocity distribution involved in the calculation of $\kappa_\parallel$ and $\beta_\parallel$ depend on the value of $R$; and (2) the moments involved in the calculation of $\kappa_\parallel$ are generally higher than those involved in the calculation of $\beta_\parallel$, it follows that the signs of the various thermoelectric coefficients are not necessarily invariant with respect to the value of $R$.

Consider for definiteness the case of a negative temperature gradient, i.e., a temperature that decreases in the positive $z$ direction (the case of a positive temperature gradient is entirely similar).  Also consider first the collisional (low-$R$) limit, for which the fourth-power dependence of the collision frequency $\nu$ causes all the moments involved to be quite high, specifically, the ninth and eleventh moments for $\kappa_\parallel$ (Equation~(\ref{kappa-parallel})) and the seventh and ninth moments for $\beta_\parallel$ (Equation~(\ref{beta-parallel})).  These high velocity moments are all dominated by large values of $v$; thus, as we move toward increasing values of $z$ (i.e., toward regions of lower temperature), the predominant effect is a reduction in $f_0(z,v)$ with $z$: $\partial f_0(z,v)/\partial z < 0$.  Both the heat flux and electron flow are predominantly carried by high-velocity electrons that move in the positive-$z$ direction to fill this relative void in $f_0$, and hence they are both oriented in the positive-$z$ direction, i.e., along the direction of decreasing temperature. Indeed, we see from Equation~(\ref{f1-general}) that the lead anisotropic term is aligned with the positive-$z$ direction: $f_1(z,v,\mu) \propto \mu$.  The coefficients $\kappa_\parallel$ and $\beta_\parallel$ are therefore both positive (the former since $q_\parallel \sim -\kappa_\parallel \, dT_e/dz$ and the latter since the conventional current $j_\parallel \sim \beta_\parallel \, dT_e/dz$ flows in the opposite direction to the electron flow, i.e., in the negative-$z$ direction).

However, in the turbulence-dominated (high-$R$) limit, the velocity-distribution moments involved are all (four powers) lower because of the weaker velocity dependence of the collision frequency $\nu = \nu_T$ (compare Equation~(\ref{alpha-def}) [with $\alpha = 0$] and Equation~(\ref{lambdac-def})).  Since the expression~(\ref{kappa-parallel}) for $\kappa_\parallel$ still involves relatively-high-order velocity moments (fifth and ninth), this reduction of all the moment orders is not sufficient to reverse the direction of the heat flux; it remains parallel to the direction of the negative temperature gradient.  However, the considerably lower-order velocity moments (third and fifth) present in the expression~(\ref{beta-parallel}) for $\beta_\parallel$ are such that the value of the pertinent moments are now dominated by lower-velocity electrons. At these velocities a decrease in temperature causes an {\it increase} in $f_0(z,v)$.  At the pertinent velocities, the spatial derivative of the isotropic zero-order term is therefore now positive: $\partial f_0(z,v)/\partial z > 0$, and so (Equation~(\ref{f1-general})) the lead anisotropic term now aligns in the negative-$z$ direction ($f_1(z,v,\mu) \propto - \mu$).  The net result is that the predominant (low-velocity) electron flow is now in the negative-$z$ direction, i.e., in the direction of positive temperature gradient, antiparallel to the heat flux.  The conventional current and the heat flux are now both in the positive $z$ direction, and so $\kappa_\parallel > 0$ and $\beta_\parallel < 0$.

We remind the reader that all the above results have been derived under the simplifying assumption that the turbulent collision frequency $\nu_T$ is independent of pitch angle $\mu$, corresponding to isotropic scattering.  While this has provided illustrative results on the impact of turbulent scattering on the various transport coefficients, in actual physical situations the scattering rate $\nu$ may depend on the pitch angle $\mu$ (and/or on the velocity $v$). This is indeed the case when angular scattering is produced by the presence of a spectrum of magnetic fluctuations in the plasma, next to be considered.

\section{Transport reduction from pitch-angle scattering by transverse magnetic fluctuations}\label{dependent}

Magnetostatic fluctuations $\delta \mathbf{B}_\perp$ perpendicular to the background magnetic field $B_0 \hat{\mathbf{z}}$ are a source of pitch-angle scattering in magnetized plasmas.  For slab turbulence the pitch-angle diffusion coefficient takes the form

\begin{equation}\label{stu-w}
D_{\mu\mu}^{T} = \frac{\pi}{2} (1-\mu^2) \, \Omega_{\rm ce} \,
\left . \frac{k_\parallel \, W(k_\parallel)}{B_0^2} \, \right \vert _{k_\parallel = \Omega_{\rm ce}/v_\parallel} \, \,\,\, ,
\end{equation}
where $W(k_\parallel)$ is the spectral energy density of the magnetic fluctuations in wavenumber, i.e., $\int W(k_\parallel) \, dk_\parallel = (\delta B_\perp)^2$. The significance of the condition

\begin{equation}\label{omegace}
\Omega_{\rm ce} = k_\parallel v_\parallel \equiv k_{\parallel} \, |\mu| \, v
\end{equation}
in the expression~(\ref{stu-w}) is that scattering of electrons occurs as a result of gyro-resonance with zero-frequency magnetic modes during their cyclotronic orbits.

We shall consider the following form of the wavenumber spectrum:

\begin{equation}
W(k_{\parallel})=C(q) \, (\delta B_{\perp})^{2} \, \lambda_{B} \, \left [ 1+(k_\parallel \lambda_{B})^{2} \right ]^{-q} \,\,\, ,
\end{equation}
where $C(q)$ is a normalization constant.  Substituting this in Equation~(\ref{stu-w}) and using Equation~(\ref{dmumudef}) gives the corresponding turbulent scattering frequency:

\begin{equation}
\nu_{T} = \frac{2D^T_{\mu\mu}}{1-\mu^2} = \pi \, C(q) \, \Omega_{\rm ce} \,
\left(\frac{\delta B_{\perp}}{B_{0}}\right)^{2} \, \left(\frac{\Omega_{ce} \, \lambda_{B}}{|\mu| v}\right) \left[1+\left(\frac{\Omega_{ce}\lambda_{B}}{|\mu| v}\right)^{2}\right]^{-q} \,\,\, .
\end{equation}
Notice the explicit dependence on $\mu$ which originates from the resonance condition~(\ref{omegace}). Introducing the normalized gyroradius (or rigidity parameter)

\begin{equation}
r_{\rm te} = \frac{v_{\rm te}}{\Omega_{\rm ce}\lambda_{B}} \,\,\, ,
\end{equation}
we obtain, in the regime $r\ll 1$,

\begin{equation}
\nu_{T} = \pi \, C(q) \, \Omega_{\rm ce} \left( \frac{\delta B_{\perp}}{B_{0}} \right)^{2}
r_{\rm te}^{2q-1} \left (\frac{v}{v_{\rm te}} \right )^{2q-1} \, |\mu|^{2q-1} \,\,\, ,
\end{equation}
which has the form (cf. Equation~(\ref{nu_t_anisotropic}))

\begin{equation}\label{dmumu-t-mu}
\nu_{T} = \left (\frac{v_{\rm te}}{\lambda_{0}} \right ) \left (\frac{v}{v_{\rm te}} \right )^{1-\alpha} \, |\mu|^{\beta} \,\,\, ,
\end{equation}
with

\begin{equation}\label{alpha-beta-q}
1-\alpha=\beta=2q-1
\end{equation}
and

\begin{equation}\label{lamb}
\lambda_{0} =\frac{v_{\rm te}}{\pi \, C(q) \, \Omega_{\rm ce} \left( \frac{\delta B_{\perp}}{B_{0}} \right)^{2}
r_{\rm te}^{2q-1}}=\frac{1}{\pi \, C(q)} \, \lambda_{B} \, \left ( \frac{B_{0}}{\delta B_\perp} \right )^{2} \, r_{\rm te}^{2-2q}\,\,\, .
\end{equation}
Note that the parameter $\lambda_{0}$ is \emph{not} the turbulent mean free path.  Rather, the turbulent mean free path is defined in terms of the parallel diffusion coefficient, which is itself
a functional of the pitch-angle diffusion coefficient (or the scattering frequency):

\begin{equation}
\lambda_{T}\equiv \frac{3}{v} \, D_{\parallel}=\frac{3v}{8}\int _{-1}^{+1}d\mu \, \frac{(1-\mu^{2})^2}{D_{\mu\mu}^{T}}=
\frac{3v}{2}\int _{0}^{1}d\mu \, \frac{(1-\mu^{2})}{\nu_{T}}\,\,\, .
\end{equation}
Substituting for $\nu_{T}$ from Equation~(\ref{dmumu-t-mu}), we obtain an expression for the turbulent mean free path:

\begin{equation}
\lambda_T = \lambda_{T0} \left (  \frac{v}{v_{\rm te} } \right)^{\alpha} = \lambda_{T0} \left (  \frac{v}{v_{\rm te} } \right)^{2 - 2q} \,\,\, ,
\end{equation}
where $\lambda_{T0}$ is the turbulent mean free path at $v=v_{\rm te}$:

\begin{eqnarray}\label{lambda_T}
\lambda_{T0} &=& \frac{3}{2 \pi \, C(q)} \, \lambda_{B} \, \left(\frac{B_{0}}{\delta B_{\perp}}\right)^{2} r_{\rm te}^{2-2q}\int _{0}^{1}d\mu \, (1-\mu^{2}) \, \mu^{1-2q} \cr
&=&\frac{3}{4 \pi \, C(q) (1-q)(2-q)} \,
\lambda_{B} \left(\frac{B_{0}}{\delta B_{\perp}}\right)^{2} r_{\rm te}^{2-2q} \,\,\, .
\end{eqnarray}

For a typical spectrum, e.g., $2q=5/3$, $\lambda_{\rm T}\sim v^{2-2q}\sim v^{1/3}$, so that $\lambda_{\rm T}$ is only weakly dependent on velocity and hence on temperature ($\lambda_T \propto T_{e}^{1/6}$). Therefore, the case of a velocity-independent (or temperature-independent) mean free path (which corresponds precisely to the case $q=1$) deserves particular attention. However Equation~(\ref{lambda_T}) shows that the turbulent mean free path formally diverges as $q \rightarrow 1$, which seems to suggest (Equation~\ref{lambda-eff-c-t})) that there is no reduction of transport below the collisional value in this case. In practice this divergence in $\lambda_T$ is avoided by considering the additional influence of collisions, with the result that the turbulent mean free path becomes finite and weakly (logarithmically) dependent on velocity \citep[see below and the appendix in][]{2014ApJ...780..176K}.

Returning to the result~(\ref{df1dmu}), which for an isobaric plasma can be written (cf. Equation~(\ref{f1-basic}))

\begin{equation}\label{df1bdmu}
\frac{\partial f_1}{\partial \mu} = - \frac {v}{\nu (v)} \left [ \left ( \frac{m_e v^2}{2k_B T_e}-\frac{5}{2} \right )
\frac{1}{T_e} \, \frac{dT_e}{dz}+\frac{eE_\parallel}{k_B T_e} \right ] \, f_0 \,\,\, ,
\end{equation}
we now have

\begin{equation}\label{dmumu-r-mu-x}
\nu(v) = \frac{v_{\rm te}}{\lambda_{\rm ei}} \, \frac{1+R|\mu|^{2q-1}x^{2q+2}}{x^3} \,\,\, ,
\end{equation}
where we have used the same normalization $x=v/v_{\rm te}$ and definition of the ratio $R = \lambda_{\rm ei}/\lambda_{0}$ as before. Substituting Equation~(\ref{dmumu-r-mu-x}) into Equation~(\ref{df1bdmu}), we obtain

\begin{equation}\label{df1bdu-r-mu-x}
\frac{\partial f_{1}}{\partial \mu} = - \lambda_{\rm ei} \, \frac{x^4}{1+R|\mu|^{2q-1}x^{2q+2}} \, \left [ \left ( x^2-\frac{5}{2} \right )
\frac{1}{T_e} \, \frac{dT_e}{dz} + \frac{eE_\parallel}{k_B T_e} \right ] \, f_0 \,\,\, .
\end{equation}
This is the generalization of Equation~(\ref{f1-x-mu}) to the case where $D^{T}_{\mu\mu}$ is proportional to $|\mu|^{2q-1}$ (Equations~(\ref{dmumu-t-mu}) and~(\ref{alpha-beta-q})).

Since, in the absence of collisions, the turbulent mean free path diverges for $q \geq 1$, the transport coefficients will also diverge for $q\geq1$ in the absence of collisions. We therefore focus on the physically-relevant $q < 1$ ($\alpha > 0$) cases.  The singular case $q=1$ ($\alpha =0$) will be dealt with separately.

\subsection{The case $q < 1$}\label{smallq}

Except for the additional $\mu$ integral which produces a factor that depends on the value of the dimensionless spectral index $q$, there will be no difference in the scalings for large $R$ compared to those already given above for the case of isotropic scattering. Hence, while the necessary integrals can be evaluated in the large $R$ limit, it is simpler to replace the anisotropic scattering problem by an equivalent isotropic one, i.e., set

\begin{equation}\label{nu-isotropic}
\nu(v) = \frac{v_{\rm te}}{\lambda_{\rm ei}} \, \frac{1+R x^{2q+2}}{x^3} \,\,\, ,
\end{equation}
but with

\begin{equation}
R=\frac{\lambda_{\rm ei}}{\lambda_{T0}}
\end{equation}
instead of $R=\lambda_{\rm ei}/{\lambda_{0}}$. This will lead to the previously given results of Section~\ref{modified-spitzer} with a somewhat different definition of $R$; these results are valid provided $q<1$, i.e., $\alpha>0$.

\subsection{The case $q=1$}\label{q1}

In the Lorentzian case $q=1$, corresponding to

\begin{equation}\label{W-spectrum}
W(k_\parallel) = \frac{(\delta B_\perp)^2}{\pi} \, \frac{(1/\lambda_B)}{(1/\lambda_B)^2+k_\parallel^2} \,\,\, ,
\end{equation}
we have

\begin{equation}\label{nuT-Lorentzian}
\nu_T = |\mu| \left ( \frac{\delta B_\perp}{B_0} \right )^2 \frac{v}{\lambda_B} = \frac{|\mu| v}{\lambda_0} \,\,\, ,
\end{equation}
so that

\begin{equation}\label{Rm}
R = \frac{\lambda_{\rm ei}}{\lambda_{0}} = \left ( \frac{\lambda_{\rm ei}}{\lambda_B} \right ) \,
\left ( \frac{\delta B_\perp}{B_0} \right )^2 \simeq \frac{ 10^4 \, T_e^2 \, (\delta B_\perp/B_0)^2}{n_e \, \lambda_B} \,\,\, .
\end{equation}
Equation~(\ref{df1bdu-r-mu-x}) can be analytically integrated over $\mu$ to give

\begin{equation}\label{f1-r-mu-x}
f_1 = \begin{cases}
- \!\!\!\!\! &\frac{\lambda_{\rm ei}}{R} \, \ln (1+Rx^4\mu) \, \left [ \left ( x^2-\frac{5}{2} \right ) \frac{1}{T_e} \, \frac{dT_e}{dz} +
\frac{eE_\parallel}{k_B T_e} \right ] \, f_0 \quad ; \quad \mu > 0 \cr
& \frac{\lambda_{\rm ei}}{R} \, \ln (1- Rx^4 \mu)\, \left [ \left ( x^2-\frac{5}{2} \right ) \frac{1}{T_e} \frac{dT_e}{dz} +
\frac{eE_\parallel}{k_B T_e} \right ] \, f_0 \quad ; \quad \mu < 0 \,\,\, .
\end{cases}
\end{equation}

This result can now be substituted in the expressions for the normalized heat flux (cf. Equation~(\ref{qparallel-mu-v-f1})) and the current density (Equation~(\ref{jparallel-mu-v-f1})).  Exploiting the hemispherical antisymmetry of Equation~(\ref{f1-r-mu-x}) in evaluating the integral over $\mu$, we obtain

\begin{equation}\label{qparallel-x-mu-r-01}
q_{\parallel} = - \, \frac{4 n_e k_B T_e v_{\rm Te} \lambda_{\rm ei}}{\sqrt{\pi} \, R} \int_0^\infty dx \, x^5 \left [ \left ( x^2-\frac{5}{2} \right )
\frac{1}{T_e} \, \frac{dT_e}{dz}+\frac{eE_\parallel}{k_B T_e} \right ] \, e^{-x^2} \, \int_0^1 d\mu \, \mu \ln ( 1+Rx^4 \mu )  \,\,\,
\end{equation}
and

\begin{equation}\label{jparallel-x-mu-r-01}
j_{\parallel} = \frac{4 n_e \, e \, v_{\rm Te} \lambda_{\rm ei}}{\sqrt{\pi} \, R} \int_0^\infty dx \, x^3 \left [ \left ( x^2-\frac{5}{2} \right )
\frac{1}{T_e} \, \frac{dT_e}{dz} + \frac{eE_\parallel}{k_B T_e} \right ] \, e^{-x^2} \, \int_0^1 d\mu \, \mu \, \ln ( 1+Rx^4 \mu )  \,\,\, .
\end{equation}
Making the substitution $y=1+Rx^4 \mu$ and then integrating by parts, the integral over $\mu$ evaluates to

\begin{equation}\label{integral-evaluation}
\int_0^1 d\mu \, \mu \ln ( 1+Rx^4 \mu ) = \frac{1}{2 (Rx^4)^2} \left \{  \left [ (Rx^4)^2 - 1 \right ] \ln (1 + Rx^4) + Rx^4 \left ( 1 - \frac{1}{2} Rx^4 \right ) \right \} \,\,\, .
\end{equation}

The values of the thermoelectric transport coefficients $\kappa_\parallel, \alpha_\parallel$, $\beta_\parallel$ and $\sigma_\parallel$ may now be obtained by substituting the expression~(\ref{integral-evaluation}) for the $\mu$ integral in the expressions~(\ref{qparallel-x-mu-r-01}) and~(\ref{jparallel-x-mu-r-01}).  These transport coefficients are best represented in relation to their Spitzer ($R=0$) values. We find

\begin{equation}\label{kappa-ratio-general}
\frac{\kappa_{\parallel}}{\kappa_{\parallel, S}} = \frac{3}{2R} \, \, \frac{\int_0^\infty \frac{1}{(Rx^4)^2} \left \{  \left [ (Rx^4)^2 - 1 \right ] \ln (1 + Rx^4) + Rx^4 \left ( 1 - \frac{1}{2} Rx^4 \right ) \right \} \, x^5 \, \left ( x^2-\frac{5}{2} \right ) \, e^{-x^2} \, dx}{\int_0^\infty x^9 \, \left ( x^2 - \frac{5}{2} \right ) \, e^{-x^2} \, dx} \,\,\, ;
\end{equation}

\begin{equation}\label{alpha-ratio-general}
\frac{\alpha_{\parallel}}{\alpha_{\parallel,S}} = \frac{3}{2R} \, \frac{\int_0^\infty \frac{1}{(Rx^4)^2} \left \{  \left [ (Rx^4)^2 - 1 \right ] \ln (1 + Rx^4) + Rx^4 \left ( 1 - \frac{1}{2} Rx^4 \right ) \right \} \, x^5 \, e^{-x^2} \, dx}{\int_0^\infty x^9 \, e^{-x^2} \, dx} \,\,\, ;
\end{equation}

\begin{equation}\label{beta-ratio-general}
\frac{\beta_{\parallel}}{\beta_{\parallel,S}} = \frac{3}{2R} \, \frac{\int_0^\infty \frac{1}{(Rx^4)^2} \left \{  \left [ (Rx^4)^2 - 1 \right ] \ln (1 + Rx^4) + Rx^4 \left ( 1 - \frac{1}{2} Rx^4 \right ) \right \} \, x^3 \, \left ( x^2-\frac{5}{2} \right ) \, e^{-x^2} \, dx}{\int_0^\infty x^7 \, \left ( x^2-\frac{5}{2} \right ) \, e^{-x^2} \, dx} \,\,\, ;
\end{equation}
and

\begin{equation}\label{sigma-ratio-general}
\frac{\sigma_{\parallel}}{\sigma_{\parallel,S}} = \frac{3}{2R} \, \frac{\int_0^\infty \frac{1}{(Rx^4)^2} \left \{  \left [ (Rx^4)^2 - 1 \right ] \ln (1 + Rx^4) + Rx^4 \left ( 1 - \frac{1}{2} Rx^4 \right ) \right \} \, x^3 \, e^{-x^2} \, dx}{\int_0^\infty x^7 \, e^{-x^2} \, dx} \,\,\, .
\end{equation}

For $R \ll 1$, we have a collision-dominated regime, and the results~(\ref{kappa-ratio-general}) through~(\ref{sigma-ratio-general}) should approach unity.  Indeed, for $R \ll 1$, the logarithm may be expanded in a Taylor series $\ln (1+q) = q - q^2/2 + q^3/3 - \ldots$, giving

\begin{equation}\label{int-r-small}
\frac{1}{2 (Rx^4)^2} \left \{  \left [ (Rx^4)^2 - 1 \right ] \ln (1 + Rx^4) + Rx^4 \left ( 1 - \frac{1}{2} Rx^4 \right ) \right \} \, {\overset{R \ll 1}\longrightarrow} \,\, \frac{1}{3} \, R x^4
\end{equation}
in place of Equation~(\ref{integral-evaluation}).  (This result may also be obtained quickly by expanding the logarithm to first order in the integral $\int_0^1 d\mu \, \mu \ln ( 1+Rx^4 \mu )$.) Using the limiting form in the numerators of~(\ref{kappa-ratio-general}) through~(\ref{sigma-ratio-general}) yields unity for all four ratios, as it should.

On the other hand, for the turbulence-dominated transport regime characterized by $R \gg 1$, we obtain

\begin{equation}\label{kappa-ratio-rggg1}
\frac{\kappa_{\parallel}}{\kappa_{\parallel, S}} \, {\overset{R \gg 1}\longrightarrow} \,\, \frac{3 \int_0^\infty \, x^5 \,  \left ( x^2-\frac{5}{2} \right ) \,  e^{-x^{2}} \, dx }{2 \int_0^\infty x^9 \, \left ( x^2-\frac{5}{2} \right ) \, e^{-x^2} \, dx}  \, \left ( \frac{\ln R}{R} \right ) = \frac{1}{40} \, \left ( \frac{\ln R}{R} \right ) \,\,\, ;
\end{equation}

\begin{equation}\label{alpha-ratio-rggg1}
\frac{\alpha_{\parallel}}{\alpha_{\parallel, S}} \, {\overset{R \gg 1}\longrightarrow} \,\, \frac{3 \int_0^\infty \, x^5 \,  e^{-x^{2}} \, dx }{2 \int_0^\infty x^9 \, e^{-x^2} \, dx} \, \left ( \frac{\ln R}{R} \right ) = \frac{1}{8} \, \left ( \frac{\ln R}{R} \right ) \,\,\, ;
\end{equation}

\begin{equation}\label{beta-ratio-rggg1}
\frac{\beta_{\parallel}}{\beta_{\parallel, S}} \, {\overset{R \gg 1}\longrightarrow} \,\,  \frac{3 \int_0^\infty \, x^3 \,  \left ( x^2-\frac{5}{2} \right ) \,  e^{-x^{2}} \, dx }{2 \int_0^\infty x^7 \, \left ( x^2-\frac{5}{2} \right ) \, e^{-x^2} \, dx} \, \left ( \frac{\ln R}{R} \right ) = - \frac{1}{12} \, \left ( \frac{\ln R}{R} \right ) \,\,\, ;
\end{equation}
and

\begin{equation}\label{sigma-ratio-rggg1}
\frac{\sigma_{\parallel}}{\sigma_{\parallel, S}} \, {\overset{R \gg 1}\longrightarrow} \,\, \frac{3 \int_0^\infty \, x^3 \,  e^{-x^{2}} \, dx }{2 \int_0^\infty x^7 \, e^{-x^2} \, dx} \, \left ( \frac{\ln R}{R} \right ) = \frac{1}{4} \, \left ( \frac{\ln R}{R} \right ) \,\,\, .
\end{equation}
Note both the $\ln R/R$ dependence of these expressions and also the change of sign of $\beta_\parallel$ at large $R$ (which occurs for the same reasons given in the discussion at the end of Section~\ref{modified-spitzer}).

\section{Application to solar flares}\label{application}

\subsection{Heating and cooling of flare coronal plasma}\label{coronal-heating-cooling}

A solar flare is ubiquitously characterized by enhanced emission in soft X-rays \citep[e.g.,][]{1969MNRAS.144..375C,1970MNRAS.151..141C,1982SoPh...78..107A}; indeed, the commonly-used GOES classification of flares is based on the soft X-ray flux in the $(1-8)$\AA~soft X-ray waveband.  The plasma responsible for emitting these soft X-rays is generally located in the corona and has a temperature $\gapprox 10^7$~K \citep[e.g.,][]{2011ApJ...727L..52R}.  The {\it in situ} heating \citep{1983ApJ...266..383C,2004ApJ...609..439F,2007ApJ...659..750R} to these temperatures is generally believed to be due to a combination of ohmic heating associated with current dissipation in the primary site of magnetic reconnection, plus collisional heating by accelerated nonthermal particles, notably electrons.  Ohmic heating by passage of the beam-neutralizing return current (see Section~\ref{return-current}) through the flaring corona may also play a role.

The heat energy transported by both nonthermal electrons and thermal conduction \citep{2003LNP...612..161H} to the chromosphere causes local heating and a corresponding increase in the emission measure of $10^7$~K gas, enhancing the overall soft X-ray emission.  Further, the pressure gradients established by this rapid heating of chromospheric material through the region of radiative instability from $T\simeq 10^5$~K to $T \simeq 10^7$~K \citep{1969ApJ...157.1157C}, drive a significant hydrodynamic response of the solar atmosphere \citep[e.g.,][]{1984ApJ...279..896N,1989ApJ...341.1067M,2005ApJ...630..573A}, in particular an upward motion of soft-X-ray-emitting plasma into the corona, a process somewhat incorrectly, but nevertheless ubiquitously, termed ``chromospheric evaporation.''

The hot $10^7$~K plasma thus created subsequently cools through radiation, conduction to the chromosphere, and flow of enthalpy.  Estimating the pertinent cooling times shows that, under the assumption of classical, collisional \citep{1962pfig.book.....S} conductivity, thermal conduction generally dominates the cooling time at the highest temperatures. However, in many events the thermal plasma is sustained well beyond the duration of the impulsive hard X-ray burst (and hence heating by nonthermal electrons), for times much longer than the conductive cooling time. This has led to the suggestion \citep[e.g.,][]{1980sfsl.work..341M,2012ApJ...759...71E} that energy is somehow injected, by unknown processes, into the corona after the impulsive phase has ceased.  However, it is important to realize that this conclusion is based on thermal conductive losses governed by a purely collisional model of heat transport, which may not be valid if, as suggested above, turbulent processes play a significant role in the region of electron transport.

To illustrate, let us consider a flare volume of $V \simeq 10^{27}$~cm$^3$, with a plasma of density $n_e = 10^{10}$~cm$^{-3}$ embedded in a magnetic field $B \simeq 10^3$~Gauss. The magnetic energy density is $B^{2}/8\pi \simeq 10^{5}$~erg~cm$^{-3}$ and the total available magnetic energy is $(B^{2}/8\pi) V \simeq 10^{32}$~erg, which is a good estimate of the amount of energy released by magnetic reconnection in a large flare \citep[e.g.,][]{2012ApJ...759...71E}. Taking the flare duration as $\tau \simeq 10^2$~s, this corresponds to a released power of $10^{30}$~erg~s$^{-1}$. This energy is transformed into kinetic energy of the particles, thermal and non thermal. A fraction, say $10\%$, of this energy goes into the kinetic energy of non-thermal electrons \citep{2012ApJ...759...71E}, giving a power $10^{29}$~erg~s$^{-1}$ in accelerated electrons, a number close to that inferred from observations of hard X-ray emission in flares \citep[e.g.,][]{2011SSRv..159..107H}. This corresponds to a volumetric heating rate due to fast electrons $Q \simeq 100$~erg~cm$^{-3}$~s$^{-1}$.

With these parameters established, we can now consider the heating (by fast electrons) and cooling (by conduction) of the soft X-ray emitting plasma, governed by the energy equation

\begin{equation}\label{energy-balance}
\frac{\partial (n_e k_B T_e)}{\partial t} = - \frac{\partial q_\parallel}{\partial z} + Q \,\,\, .
\end{equation}
As we have seen above, the conductive heat flux takes the form

\begin{equation}\label{qparallel-energy}
q_{\parallel} = - \frac{2 n_e k_B \, (2 k_B T_e)^{1/2}}{m_e^{1/2}} \, \lambda \, \frac{dT_e}{dz} \,\,\, ,
\end{equation}
where $\lambda$ is the mean free path, representing the combined effect of Coulomb collisions and turbulent scattering,

\begin{equation}
\lambda= \frac{\lambda_{\rm ei}}{1+R} \,\,\,,
\end{equation}
and where the turbulent reduction factor

\begin{equation}
R =\frac{ \lambda_{ei}}{\lambda_0}=\frac{ 10^4 \, T_e^2 \, (\delta B_\perp/B_0)^2}{n_e \, \lambda_B} \,\,\, .
\end{equation}
When $R\gg1$, i.e., when the turbulent parameter $\lambda_{0}$ is much smaller than the collisional mean free path $\lambda_{\rm ei}$ (corresponding to $ \lambda_{0}\ll 10^{4}T_{e}^{2}/n_{e}$), heat transport is regulated predominantly by turbulence. We call this a ``turbulence-dominated regime,'' although we again stress (cf. Section~\ref{scattering-parameters}) that the role of collisions in maintaining a near Maxwellian distribution of background electrons remains important. On the other hand, turbulence plays a negligible role when $R \ll 1$.

Let us first balance the terms on the right side of Equation~(\ref{energy-balance}) to obtain an estimate of the steady-state temperature in the presence of a heat source $Q$ balanced by thermal conduction:

\begin{equation}
2 n_e k_B \left ( \frac{2 k_B T_e}{m_e} \right )^{1/2} \lambda \, \left ( \frac{T_e}{L^2} \right ) \simeq Q \,\,\, ,
\end{equation}
from which follows the scaling law

\begin{equation}\label{scaling-law}
T_e \simeq \frac{m_e^{1/3}}{2 k_B} \, \left ( \frac{Q L^{2}}{n_e\lambda} \right )^{2/3} \,\,\, .
\end{equation}
The vast majority of theoretical studies concerning the response of coronal loops to heating use the collisional \citep{1962pfig.book.....S} conductivity and corresponding mean free path $\lambda_{\rm ei}$; this includes the determination of equilibrium scaling laws and temperature distribution for active region loops
\citep[e.g.,][]{1978ApJ...220..643R, 1978A&A....70....1C, 2010ApJ...714.1290M} and the modelling of evaporative cooling and enthalpy-based response to coronal heating \citep[e.g.,][]{1978ApJ...220.1137A, 1995ApJ...439.1034C,2010LRSP....7....5R,2012ApJ...758....5C}.
When the mean free path is indeed close to its collisional value $\lambda_{ei}$ (Equation~(\ref{lambda-ei})), Equation~(\ref{scaling-law}) gives the scaling law appropriate to a collisional regime of transport \citep{1978ApJ...220..643R} :

\begin{equation}\label{scaling-law-collisional}
T_e \simeq \frac{m_e^{1/7}}{2 k_B} \, \left ( 2 \pi e^4 \, \ln \Lambda \right )^{2/7} Q^{2/7} \, L^{4/7} \simeq 50 \, Q^{2/7} \, L^{4/7} \,\,\, .
\end{equation}
Note that this is independent on the density $n_e$, and that this scaling follows straightforwardly from the familiar balancing relation $Q \propto T_e^{7/2}/L^2$ appropriate to a collisional environment.  Substituting the values $Q \simeq 100$~erg~cm$^{-3}$~s$^{-1}$ and $L \simeq 10^9$~cm, we obtain

\begin{equation}\label{scaling-law-collisional-result}
T_e \simeq  3 \times 10^7 \, {\rm K} \,\,\, ,
\end{equation}
a value nicely consistent with soft X-ray emission.

On the other hand, returning to Equation~(\ref{scaling-law}) with $R\gg1$, we now obtain the fundamentally different scaling law
for the equilibrium temperature in a turbulence dominated regime:

\begin{equation}\label{scaling-law-turbulent}
T_e \simeq \frac{m_e^{1/3}}{2 k_B} \, \left ( \frac{Q}{n_e} \right )^{2/3}L^{4/3}\lambda_{B}^{-2/3}
\left (\frac{\delta B_{\perp}}{B_{0}} \right)^{4/3} \,\,\, ,
\end{equation}
which now depends on the density $n_e$. With a magnetic field perturbation ratio $\delta B_\perp /B_0 \simeq 0.1$ and a magnetic correlation length $\lambda_B \simeq 10^6$~cm ($\lambda_0 \simeq 10^8$~cm) gives the significantly larger temperature

\begin{equation}\label{scaling-law-turbulence-result}
T_e \simeq  1 \times 10^8 \, {\rm K} \,\,\, .
\end{equation}
Hence, for a given heating rate $Q$ and loop properties $n_e$ and $L$, the turbulent suppression of heat transport associated with large values of $R$ leads to a higher steady-state temperature than that obtained by using Spitzer conductivity.  Plasma at temperatures $\sim$$10^8$~K ($\simeq 10$~keV) can make a meaningful contribution to the {\it hard} X-ray emission from the flare.  Since thermal conduction is inhibited,
possibly also by other collective plasma processes \citep{1979ApJ...228..592B,1980ApJ...242..799S},
such $10^8$~K temperatures will likely be confined rather than extending along the entire magnetic loop. For a given heating rate $Q$, Equations~(\ref{scaling-law-collisional}) and~(\ref{scaling-law-turbulent}) give quite different dependencies of the flaring coronal temperature $T_e$ on the loop length $L$ ($L^{4/7}$ and $L^{4/3}$, respectively), a result that should be observationally testable.

We note that the steady state temperature $T_{e}=3\times 10^{7}$~K in Equation~(\ref{scaling-law-collisional-result}) above was obtained by assuming a collisional transport regime.  Such an assumption is valid only under the dual conditions that both the Knudsen number ${\rm Kn} = \lambda_{\rm ei}/L_T$ and the turbulent reduction factor $R$ are $\ll 1$.  Consistency therefore demands that we evaluate the validity of these two assumptions.  We find that at these temperatures (and densities $n_e \sim 10^{10}$~cm$^{-3}$), the collisional mean free path $\lambda_{\rm ei}$ is actually of the order of the loop length $L \simeq 10^9$~cm; thus the Knudsen number is of order unity and so the use of a collision-related expression for the heat flux is somewhat questionable.  In such conditions the heat flux is instead determined \citep[e.g..][]{1979ApJ...228..592B} by a flux-limited value equal to a fraction of the free-streaming limit (Equation~(\ref{qparallel})), leading to a generally higher equilibrium temperature and conductive cooling time than is appropriate for the collisional case.  With regard to the second assumption, the main thrust of the present paper is that, whether or not the Knudsen number is small, the heat flux may be further limited (by a factor $\simeq R$) by the presence of collisionless pitch-angle scattering.  Thus, if the turbulent mean free path is substantially less than the collisional mean free path (or, for Knudsen numbers of order unity or more, the length of the flaring loop), then the much larger temperatures ($T_{e}\sim 10^{8}$~K; Equation~(\ref{scaling-law-turbulence-result})) appropriate to a turbulence-dominated regime apply.

We now turn to a consideration of the role of conduction in cooling the coronal plasma after the energy input $Q$ has ceased. Balancing the heating and conductive cooling in Equation~(\ref{energy-balance})), and using the expression~(\ref{qparallel-energy}), we obtain an expression for the cooling time-scale $\tau_{c}$:

\begin{equation}\label{cooling-time}
\tau_{c} \simeq \frac{m_e^{1/2}}{(2 k_B T_e)^{1/2}} \, \frac{L^2}{\lambda} =
\left ( \frac{L}{v_{\rm te}} \right ) \, \left ( \frac{L}{\lambda} \right ) = \left ( \frac{L}{v_{\rm te}} \right ) \, \left ( \frac{L}{\lambda_{\rm ei}} \right ) \, (1+R) \,\,\, .
\end{equation}
The cooling time $\tau_{c}$ is thus the free-streaming transport time-scale $L/v_{\rm te}$ of thermal electrons multiplied by the inverse of the Knudsen number ${\rm Kn} = \lambda/L$.  For a temperature $T_{e} \simeq 3 \times 10^7$~K (Equation~(\ref{scaling-law-collisional-result})), the electron thermal speed $v_{\rm Te} \simeq 3 \times 10^9$~cm~s$^{-1}$, so that for a loop length $L \simeq 10^9$~cm, the free-streaming escape time is $L/v_{\rm te} \simeq$~0.3~s.  The collisional mean free path $\lambda_{\rm ei}\simeq 10^9$~cm and hence $L/\lambda_{ei}\sim 1$, which when $R \ll 1$, yields a cooling time of the order of the free-streaming time scale. A smaller initial temperature $T_{e} \simeq  10^7$~K has the collisional mean free path decreased by an order of magnitude giving a cooling time $\tau_{c}\sim 3s$. These time scales are much shorter than the duration of the soft X-ray emission, which has led to the realization that the coronal plasma will cool very rapidly after the cessation of the energy input term $Q$, to the point where some form of post-impulsive-phase energy input to the corona is needed to sustain the soft X-ray emission for the observed times $\gapprox 100$~s \citep[see, e.g.,][]{1980sfsl.work..341M,2012ApJ...759...71E}. However, in the presence of turbulence the cooling time $\tau_{\rm c}$ is further enhanced by a factor $(1+R)$, and so cooling by thermal conduction parallel to the guiding magnetic field can thus be significantly inhibited.  This increase in the cooling time significantly reduces the previously-assumed requirement \citep{1980sfsl.work..341M,2012ApJ...759...71E} for post-impulsive phase heating of the coronal plasma.

Before moving on, we parenthetically note that similar issues regarding thermal conductivity (and its suppression) in stochastic magnetic fields arise in the context of galaxy cluster formation and in the theory of cooling flows \citep{1994ARA&A..32..277F, 1998PhRvL..80.3077C, 2001ApJ...549..402M}.

\subsection{Return current effects}\label{return-current}

The very significant electrical currents associated with the injection of non-thermal electrons necessitate a current-neutralizing return current, set up by a combination of electrostatic and inductive processes, the relative role of which has been a matter of some debate \citep{1977ApJ...218..306K,1980ApJ...235.1055E,1984ApJ...280..448S,1985ApJ...293..584H,1989SoPh..120..343L,1990A&A...234..496V,1995A&A...304..284Z,2005A&A...432.1033Z,2006ApJ...651..553Z,2008A&A...487..337B,2013ApJ...773..121C}.
However, irrespective of the detailed physics responsible for establishing this return current, driving it through the finite resistivity of the ambient medium requires a local electric field ${\cal E}_\parallel = j_\parallel/\sigma_\parallel$, which in turn causes both an Ohmic heating rate $Q_{\rm rc} = j_\parallel \, {\cal E}_\parallel = j_\parallel^2/\sigma_\parallel$ and an additional energy loss rate $| dE/dt | = e \, {\cal E}_\parallel \, v = j_\parallel \, {\cal E}_\parallel/n_e = j_\parallel^2/n_e \, \sigma_\parallel$ for each of the accelerated electrons.

Now, the transport of {\it non-thermal} electrons is dominated by non-diffusive cold-target energy losses \citep[see, e.g.,][]{1971SoPh...18..489B,1973SoPh...31..143B,1978ApJ...224..241E}, and hence we expect that the transport of such electrons, and hence the current density $j_\parallel$ that they carry, is largely unaffected by collisionless pitch-angle scattering. (Although \citet{2014ApJ...780..176K} have shown that the direct beam current $j_\parallel$ is also reduced somewhat due to the presence of pitch-angle scattering, the reduction factor is not as large as the transport coefficient reduction factors $R$ considered here, so that we may assume that the current density $j_\parallel$ associated with the injected electrons is essentially the same as in the purely collisional case.)  Any change in ohmic energy losses is therefore driven primarily by changes in the parallel electrical conductivity $\sigma_\parallel$.  Reducing the value of $\sigma_\parallel$ through turbulence results in a greater rate of Ohmic heating $Q_{\rm rc}$ (and hence higher coronal heating rates) and also a greater energy loss rate $|dE/dt|$ for the accelerated electrons.

This enhancement of the return-current electron energy loss rates affects the heating rate as a function of position \citep{1980ApJ...235.1055E} and hence the hydrodynamic response of the atmosphere \citep[e.g.,][]{1984ApJ...279..896N,1989ApJ...341.1067M}. It also reduces the amount of energy precipitating into the chromospheric footpoints, thus possibly accounting for the ``gentle'' evaporation observed by, e.g., \citet{1988ApJ...329..456Z}. Enhanced return current energy losses also result in a more effective confinement of hard-X-ray-producing electrons in the corona, which may offer an alternative explanation for loop-top coronal sources \citep{2003AdSpR..32.2489K,2008ApJ...673..576X,2012A&A...543A..53G,2012ApJ...755...32G,2014ApJ...787...86J}.

Estimates of the ratio of return current heating to collisional energy loss in the flaring corona show that, for moderately large flares they are comparable \citep[see Figure 3 of][]{1980ApJ...235.1055E}.  The same figure shows that the ratio of return current heating to collisional energy loss in the chromosphere, where most of the electron heating occurs, can be up to several percent. Thus, enhancing the return current heating/energy loss rate by even an order of magnitude through turbulent modification of the electrical conductivity $\sigma_\parallel$ and could possibly transform the flaring corona into a return-current-dominated regime \citep{1977ApJ...218..306K,1989SoPh..120..343L,1990A&A...234..496V,1995A&A...304..284Z}. This has very significant implications, ranging from the spatial distribution of hard X-ray emission and electron heating, to the total number of accelerated electrons required to produce a given hard X-ray intensity \citep{2005A&A...432.1033Z,2006ApJ...651..553Z}.

A more dominant role for return current losses in the energy loss rate for accelerated electrons has a possibly even more interesting effect. Since the energy loss rate for an individual electron $| dE/dt | = e \, {\cal E}_\parallel \, v = e \, v \, j_\parallel/\sigma_\parallel$, which is proportional to the injected current $j_\parallel$ and hence the electron injection rate, and since the total hard X-ray yield is proportional to the injection rate divided by the energy loss rate \citep{1988ApJ...331..554B}, it follows that the hard X-ray yield in a return-current-loss dominated regime {\it is independent of the injected number of electrons} \citep{1980ApJ...235.1055E}.  Such a possible saturation of hard X-ray flux with increasing flare intensity has been reported by \citet{2007ApJ...666.1268A}.

\section{Summary and conclusions}\label{summary}

Motivated by observations
\citep{2011ApJ...730L..22K, 2011A&A...535A..18B}
suggesting the presence of magnetic fluctuations in flaring loops and also \citep{2013A&A...551A.135S} suggesting that turbulent pitch-angle scattering plays a significant role in the transport of energy by both thermal and non-thermal electrons in solar flares, we have derived formulae for the thermal and electrical conductivities in the presence of both collisions and magnetic turbulence.

The enhanced electron confinement effected by the addition of collisionless pitch-angle scattering can reduce the {\it thermal} conductivity of the corona, thus decreasing thermal conductive losses and so increasing coronal temperatures compared to those in a model with collisionally-dominated transport.  This may explain the localization of coronal X-ray sources in the apex of the loop \cite[see, e.g.,][for a review]{2011SSRv..159..107H}.   It also increases the cooling time for the flare-heated coronal plasma, possibly alleviating the need for post-impulsive-phase heating by unidentified processes \citep{1980sfsl.work..341M,2012ApJ...759...71E}.  Finally, it means that the corona becomes more of a ``warm'' target in the calculation of the energy loss rate of accelerated electrons, which has an impact on the relationship between the source-integrated electron spectrum and the injected spectrum \citep{1988ApJ...331..554B,2003ApJ...595L.115B,2003ApJ...595L.119E,2011SSRv..159..301K,2015ApJ...809...35K} and hence on the overall energetics associated with accelerated electrons \citep{1986NASACP...2439..505D,1997JGR...10214631M,2003ApJ...595L..97H,2004JGRA..10910104E,2005JGRA..11011103E,2012ApJ...759...71E}.

The suppressed value of the {\it electrical} conductivity may significantly increase the importance of ohmic heating, both in the thermodynamics of the flare-heated atmosphere and in the propagation of the accelerated electrons themselves. In particular, because the inclusion of return current energy losses affects the ``bremsstrahlung efficiency'' (energy of hard X-rays produced per electron energy injected in the corona), it may significantly alter the injected electron flux required to produce given hard X-ray flux, with further attendant implications for the overall role of accelerated electrons in flare energetics.

Because of these important implications for quantitative details of the impulsive phase of solar flares and even for its overall viability \citep[see, e.g.,][]{2009A&A...508..993B}, we urge workers in the field to consider such anomalous transport effects in their modeling of particle transport, thermal conduction, and the electrodynamics of solar flares.

\acknowledgments This work is partially supported by a STFC consolidated grant. Financial support  by the European Commission through the ``Radiosun'' (PEOPLE-2011-IRSES-295272) is gratefully acknowledged. AGE was supported by grant NNX10AT78G from NASA's Goddard Space Flight Center.

\bibliographystyle{apj}
\bibliography{bian_et_al_turbulent_transport}

\end{document}